\documentstyle{article}
\begin{document}
\begin{titlepage}
\begin{flushright}    UFIFT-HEP-01-06 \\ 
\end{flushright}
\vskip 1cm

\centerline{\LARGE\bf Testing Fundamental Ideas with $\nu\,$'s}
\vskip .5cm
\centerline{\bf Pierre Ramond}
\vskip .5cm
\centerline{Institute for Fundamental Theory,  Physics Department,} 
\centerline{University of Florida, Gainesville, FL 32611, USA}
\centerline{E-mail: ramond@phys.ufl.edu}
\vskip 1cm
\centerline{\bf {Abstract}}
\vskip .5cm
Fundamental questions in neutrino physics  are likely to be resolved by upcoming experiments. Neutrino masses  require either new Higgs particles or sterile neutrinos. Their masses, couplings,  and habitat are yet to  be determined. Neutrino physics is particularly sensitive to brane ideas as the extra particles are electroweak singlets and may free-range in the bulk. We outline some of the important theoretical ideas for generating neutrino masses, and how experiments will affect their longevity.
\vfill
\begin{flushleft}
June 2001 \\
\end{flushleft}
\end{titlepage}

\section{ Historical $\nu$'s}
Many experiments have reported  anomalies in neutrino physics. These  either detect too few or too many neutrinos of one kind, or else find  an unexpected neutrino flavor. Theoretical dogma attributes these effects to  oscillations of the common neutrinos ({\it neutrino vulgaris}) into one another or into some hitherto unknown species ({\it neutrino improvisus}). The most convincing evidence for neutrino oscillations comes  from the SuperK experiment\cite{SUPERK} on atmospheric neutrinos. The LSND collaboration\cite{LSND} has also reported direct evidence for oscillations. Eversince Ray Davis's spectacular radiochemical discovery of a solar neutrino deficit\cite{DAVIS}, much has been learned, but direct evidence of oscillations of solar neutrinos still proves elusive, although not all experimental results can be explained by oscillations  among the three common neutrino species.

Present theoretical dogma interprets these results as indications that neutrino species differ by their masses, and that mixtures of mass eigenstates are produced in weak decays. As neutrinos are strictly massless in the Standard Model, these require that  new degrees of freedom  be added to the list of elementary particles. Much of what follows will catalog the possible mechanisms of neutrino masses. 

While it is not difficult to lay out the theoretical possibilities, the present state of the data makes it difficult to single out one particular theoretical path beyond the Standard Model. Fortunately, two experiments, one destined to corroborate the Los Alamos result\cite{MINIBOONE}, the other\cite{SNO} to accurately 
measure the high energy solar neutrinos and their flavors, will soon throw light on some fundamental questions: the possible existence of new neutrinos ({\it neutrino improvisus}), and the direct measurement of solar neutrino oscillations. In the meantime,  it might be sobering to  reflect on the history of neutrino and weak interaction physics. 

History indicates that all Weak Interaction experiments start out wrong. The electron spectrum in $\beta$-decay was first measured to be discrete,  the $V-A$ theory emerged from theory, not experiments, and of course there were many premature reports of neutrino oscillations. This is no surprise as these experiments always operate at the limit of existing technologies and imagination. 
It is therefore unlikely that all recent results will stand the test of time.

Before being accused of bias against experimentalists, it might be reassuring to note that theorists, even legendary ones, are not immune to these failings. Indeed, Wolfgang Pauli himself was only half right when he postulated the existence of what became later known as the neutrino.

In his famous letter of December 6, 1930, he starts with the nerdy {\it ``Dear Radioactive Ladies and Gentlemen..."}, and says {\it ``... I have hit upon a desperate remedy to save the exchange theorem of statistics and the energy theorem. ...there could exist in the nuclei electrically neutral particles...which have spin 1/2, and... do not travel with the velocity of light. The continuous $\beta$ spectrum would then become understandable..."}.  Pauli's particle solved two problems, by living   in the nucleus it produced the correct wavefunction for some nuclei, and  the continuous electron spectrum. This was before the neutron had been discovered. Pauli needed his particle to have mass because he thought it was bound to the proton through its magnetic moment. He then goes on to say {\it ``...but I do not feel secure enough to publish anything about this idea... but ... only those who wager can win..."}. It is refreshing to see that even though he introduced one particle to solve two problems, the great Pauli still felt uneasy about it. In his generation, it might have seemed to be some sort of failure. Almost a century later, theorists do not hesitate to explain perceived anomalies by introducing new particles, nay, towers of new particles.  Pauli closes the letter with ``{\it I cannot attend personally in T\"ubingen since I am indispensable here on account of a ball...}". This has got to be one of the great {\bf dates} in the history of Science! 

Soon after, Chadwick discovers a massive spin 1/2 neutral nuclear particle, the neutron (the name Pauli had chosen for his particle).  It is Fermi who perceived the correct path: the Pauli particle  is not nuclear,  it is  produced in radioactive decay,  is one of the founding blocks of his celebrated  theory, and of course acquired an Italian name. 

In subsequent years, it was common lore that neutrinos would never be detected, but 23 years later  Cowan and Reines\cite{CORE} found evidence for antineutrinos in an ingenious scintillator experiment. Alas, it took much longer (39 years!) for Reines to be recognized by the Nobel Committee.  

The study of neutrinos,   their properties and interactions have proven to be at 
the center of the most fundamental questions  of the weak interactions. Almost  after three quarters of a century, neutrinos  still appear to have magical attributes, and much is yet to be learned from their study.

 \section{ Standard Model $\nu$'s}
In the Standard Model, neutrinos have a special place because they are purely left-handed Weyl spinors, $\nu_{e,\mu,\tau}$ -- there are no right-handed neutrinos. They are part of a weak isodoublet with $I=1/2$ and $I_3=1/2$.  Lorentz invariance allows for the flavor-symmetric {\it Majorana  mass} invariant of the form $\nu_i^T\,\nu^{}_j$, where $T$ stands for transpose and $i,j$ denote the flavor.  

This mass is the component of an isotriplet with $I_3=1$. Since there are no isotriplet Higgs to couple to this combination, Standard Model neutrinos are naturally massless at tree level. At loop levels, it is possible to generate higher dimensions operators which after electroweak symmetry breaking generate neutrino mass, since one can make an isotriplet out of two Higgs, leading to the gauge invariant interaction (first considered by Weinberg) 

$$\frac{1}{M}\,\nu^T_i\vec\tau\nu^{}_j\cdot H\vec\tau H\ ,$$
where we have inserted an unknown mass $M$ to restore dimensions. 
The Standard Model respects three global symmetries, $B-L$, where $B$ is baryon number, and $L$ is the total lepton number $L=L_e+L_\mu+L_\tau$ and the relative lepton numbers $L_e-L_\mu$, $L_\mu-L_\tau$. Since the Higgs has no lepton number, this invariant violates these global symmetries, and it will not be generated in perturbation theory. Neutrino masses can be generated {\bf only by adding extra particles to the Standard Model}.    

The superK result\cite{SUPERK} on atmospheric neutrinos requires  the Standard Model be extended. Although this comes as no surprise since the Standard Model looks like so many broken shards of beautiful structures and its parameters hint at tantalizing patterns, perhaps neutrino properties will help determine its correct extension, and contribute to the archeology of the fundamental interactions.

\section{ Massive $\nu$'s}
One may start by asking  two important questions  regarding the nature of these new degrees of freedom: their gender, fermions or bosons, and in view of recent theoretical insights from string theory, their habitat: do they live in $3+1$ dimensions or range over the whole space implied by superstring theory. 

According to recent theoretical ideas\cite{HW,LYK,ADD}, only particles free of electroweak quantum numbers, such as the graviton,  are allowed to freely range ``off the wall" in all 
extra dimensions.  However, this criterion may be ill posed, as there may yet be  hitherto unknown quantum numbers that keep these particles on the brane. 

\subsection{Leptonic Higgs}
We have seen that any extension of the Standard Model that breaks the global lepton number symmetries  generates  neutrino masses.  One way\cite{JOURNEYS} to break lepton numbers is to introduce spinless Higgs particles that carry lepton number. Lepton numbers can be broken either explicitly through their interactions, or spontaneously. In the latter case, there will result the associated massless Nambu-Goldstone boson, called the Majoron, which cannot have electroweak quantum numbers. In order to carry lepton numbers, these new LeptoHiggs must couple to the  Standard Model leptons.  There are only three possible Yukawa interactions to consider

$$L_{(i}\vec\tau\,L_{j)}\cdot\vec T\ ,\qquad L_{[i}\,L_{j]}\,S^+\ ,\qquad \overline e_i\,\overline e_j\,\overline S^{--}\ ,$$
where $(...)$ and $[...]$ indicate symmetrization and antisymmetrization 
over the family indices, respectively. The isotriplet LeptoHiggs $\vec T$, and the two isosinglets, $S^+$ and $S^{++}$ each carry two units of lepton number. 

There are three possible  cubic gauge-invariants among the Higgs particles, $S^+\,S^+\,\overline S^{--}$, $\vec T\cdot\vec  T\,\overline S^{--}$, and $H\vec\tau H\cdot \vec T$, which all carry six units of lepton number. These models, where  lepton number is explicitly broken, yield Majorana neutrino masses from  loop effects, but they all depend on a new mass parameter, the strength of these cubic couplings. 

\subsection{Sterile $\nu$'s}
Sterile neutrinos are  Weyl fermions with no electroweak quantum numbers.  They can be endowed with  various lepton numbers through the couplings
$$L_i\,H\,\overline N_a\ ,$$
where $a=1,2,\dots,n_s$. The notation is such that  $N_a$  carries positive lepton numbers, and we do not know  "who ordered these fermions", their number,  their masses, and if they are free-ranging or stuck on the brane.

Since they have no apparent quantum numbers, we can add to these interactions  Majorana mass terms, without  breaking local symmetries

$$M_{(ab)}\overline N_a\,\overline N_b\ .$$
For simplicity consider only one such fermion and one family of particles, then $M_{ab}\equiv M$. We may consider two cases:  
\begin{itemize}
\item $M\ne 0$, an allowed possibility  which explicitly breaks lepton number. 
 In the spirit of effective field theories, we expect $M$ to be of the scale of the 
cut-off, since $M$ breaks no local symmetries. This will not only generate neutrino masses but also indicate through their values, where the Standard Model cut-off lies. Electroweak breaking generates the tree-level neutrino matrix

$$\pmatrix{0&m\cr m& M}\ .$$
Here $m$ is the mixing induced by the Higgs vev, and is limited to be of the order of the electroweak breaking $\sim 246$ GeV. This matrix has two eigenvalues
$\frac{1}{2}(M\pm\sqrt{M^2+4m^2})$. If $M\sim m$, then both eigenvalues are ${\cal O}(m)$, and expected to be of electroweak strength, which does not explain the great  disparity between neutrino masses and those of their associated charged leptons. However, when $M>>m$, the two eigenvalues split into a light eigenvalue

$$m_\nu\,\sim\,m\Big(\frac{m}{M}\Big)\ ,\qquad m_N\,\sim\, M\ ,$$
which explains the relative suppression\cite{SEESAW} from the scale ratio between the electroweak scale and that of M. In this sense, the smallness of the neutrino masses is a reflection of a high cut-off for the Standard Model. If   mass squared differences from the solar and atmospheric neutrino experiments reflect  the actual neutrino  masses,   a cut-off close to $10^{15}$ GeV (with errors in the exponent!) emerges, which is close to the scale at which the Supersymmetric Standard Model coupling constants unify. Since neither coupling unification nor such a large scale make sense without perturbative control, the  smallness of neutrino masses adds to the circumstantial evidence for supersymmetry. The case $M\sim 0$ is not natural unless there are quantum numbers which keep it small. However if both $M$ and $m$ are very small, neutrino masses will be tiny.

\item The case $M=0$ is natural in the sense that it preserves lepton number. This produces a Dirac neutrino mass, preserves lepton number, and $\overline N$ is truly the Dirac partner of {\it neutrino vulgaris}. Again naive expectations produce masses akin to those of the charged leptons.  
These considerations do not apply if, for some reason, the dimensionless Yukawa couplings are anomalously small. 
  
\end{itemize}

\subsection{ Off the Wall $\nu$'s}
The existence of charge-carrying branes in string and M-theory  has motivated the idea that our three-dimensional world is itself a brane embedded in a higher dimensional space\cite{HW,LYK}. Particles with brane-local gauge invariance are localized on the brane, the others are free-ranging.   

Assuming that there are no more brane-local invariances, 
we may then consider all particles with no Standard Model quantum numbers as possible free-rangers. Putative free-range particles are the graviton, the sterile neutrinos\cite{ADDM}, and possible electroweak singlet Higgs (e.g. the axion). 

In the effective theory in $3+1$ dimensions below the scale at which this structure becomes apparent (from TeVs to the Planck scale?), there will appear   anomalously small dimensionless couplings, as well as new particles of geometrical origin, the Kaluza-Klein modes.

To see how this comes about, consider spinor and scalar fields in $4+1$ dimensions. Fermions have engineering dimensions $2\,(3/2)$ and bosons  $3/2\,(1)$ in five (four) space-time dimensions, as is easily seen from their kinetic terms. Consequently, brane-bulk couplings which take place over the $3+1$ space-time have units, leading to

$$\frac{1}{\sqrt M_s}\int d^4x\,L\,H\,\overline N_5\ ,\qquad \frac{1}{\sqrt M_s}\int d^4x\,H^3\,\Phi_5\ ,$$
where $N_5,\,\Phi_5$ are free-rangers, and $L,\,H$ are localized to the brane. The mass parameter $M_s$ is set by the underlying theory. The next step is to expand the five-dimensional fields in terms of four-dimensional fields, assuming periodic boundary conditions in the extra space dimension, $y$. We can have periodic (toroidal for the sophisticate) boundary conditions 

$$
 \Phi_5(x,-\pi R)=\Phi_5(x,\pi R)\; .
$$
Correspondingly, the free-range fields can be expanded into Kaluza-Klein
modes as
$$
 \Phi_5(x,y) = \frac{1}{\sqrt{2\pi R}}\sum_{n}\Phi_{n}(x)
               \exp\left(\frac{iny}{R}\right)\; ,
$$
where the $\Phi_n(x)$ describe the four-dimensional Kaluza-Klein tower. 
Putting back in the Lagrangian, we see that the dimensionless Yukawa coupling between the brane fields and the KK tower states acquires a geometrical factor

$$\frac{1}{\sqrt{2\pi\,M_sR}}\ ,$$
which can account for their anomalously low value, if $M_s\,R>>1$. 

With an extra   spinless bulk field, $\Phi_5$, we could generalize  the LeptoHiggs models by considering the couplings

$$L_{(i}\vec\tau\,L_{j)}\cdot\vec T\frac{\Phi_5}{\sqrt M_s}\ ,\qquad L_{[i}\,L_{j]}\,S^+\frac{\Phi_5}{\sqrt M_s}\ ,\qquad \overline e_i\,\overline e_j\,\overline S^{--}\frac{\Phi_5}{\sqrt M_s}\ ,$$
which  gives the complex bulk field  six units of lepton number. To break lepton number and generate neutrino masses, we   add  terms polynomial in $\Phi_5$. If the bulk field is real, the coupling preserves discrete lepton symmetries $\exp(i\pi/3)$ which are enough to keep neutrinos massless.  

Extra bulk  neutrinos give a generic mechanism to generate  small neutrino masses. We add to the brane-bulk coupling a bulk mass $\mu$

$$ 
\frac{1}{\sqrt M_s}\int d^4x\,L\,H\,\overline N~+~\int d^5x\,\mu\overline N_5\overline N_5\ .$$
As we have seen the case $\mu=0$ generates a small mass if the geometry is such that $M_sR>>1$. On the other hand, if $\mu$ is small, it will generate a Kaluza-Klein tower of states with masses $\mu$, $\sqrt{\mu^2+\frac{1}{R^2}}$, etc... .

If for some reason, $\mu$ is small (if the $\overline N_5$ are modulini, they would be massless up to supersymmetry breaking\cite{MODULINI,OXFORD}), then the tower of sterile KK particles would be light enough to play a role in the phenomenology of mixing. In this sense, the KK pictures generates sterile neutrinos. 

If light sterile neutrinos are found necessary to explain the data,  this picture will become very appealing. Until  experiments become precise enough to challenge the conservative view of three standard model neutrinos and no {\it light} sterile neutrinos, we will continue in the belief that ``{\it a neutrino on my brane is better than two in the bulk} ".

\section{ Experimental $\nu$'s}
More data is needed before we can safely assert one theoretical extension of the Standard Model over another. Of particular interest are  three crucial queries:
1)  $\nu$-mass hierarchies; 2) 
 $\nu$-mixing patterns; 3) 
 Number of light $\nu\,$s.

Recent experiments find  anomalies in their study of neutrino transfers in three different areas:
\begin{itemize}
\item The number of $\nu_e$ coming from the core of the Sun is roughly half the expected value from solar models\cite{BAHCALL}. This fact was first detected in the highest energy solar neutrinos, coming from ${}^8B$ decay; it has since been since observed in the lowest fusion $pp$ neutrinos\cite{SAGE,GALLEX}. While there is no direct evidence for neutrino oscillations, it is the preferred theoretical interpretation. The effect is enhanced by resonant oscillation while the neutrinos travel out of the Solar core, via the MSW effect\cite{MSW}. The present data may be explained by oscillation of $\nu_e$ into another neutrino, with possible parameters: large angle mixing with $\Delta m^2$ from  $10^{-4,}~ {\rm eV}^2$ (Adiabatic MSW) 
to $10^{-10,-12}~{\rm eV}^2$ (Vacuum), and small angle with $\Delta m^2\sim 10^{-5}~{\rm eV}^2$ (NonAdiabatic MSW). Recent results\cite{SUPERKSOLAR} which find little or no distortion in the ${}^8B$ neutrino favor the large angle scenarios.  

\item The number of $\nu_\mu$ produced in cosmic ray collisions with the atmosphere is also below expectations, and its directional flux correlates with oscillations of $\nu_\mu$ into $\nu_\tau$ with large mixing angle, and $\Delta m^2\sim 3\times 10^{-3}~{\rm eV}^2$. This result is corroborated by other experiments, notably by the SOUDAN and MACRO results.

\item The LSND\cite{LSND} experiment at Los Alamos has observed an excess of $\overline\nu_e$ in a $\overline\nu_\mu$ beam, leading to a much bigger $\Delta m^2\sim 1~{\rm eV}^2$.  This experiment stands on its own, with much of its allowed parameter space already ruled out by the KARMEN\cite{KARMEN} experiment. 
\end{itemize}
It is not possible to accomodate these three sets of results with just three neutrinos. A fourth neutrino is needed, and since the $Z$-width forbids another neutrino, the fourth cannot interact weakly. This leads to a light sterile neutrino. As we have argued the theoretical implications of such a particle are immense, requiring a new principle for keeping a small mass for a fermion without apparent quantum numbers.

Fortunately, a new set of experimental results will soon become available. 
\begin{itemize}
\item The SNO detector\cite{SNO} is presently taking data on the charged current scattering of ${}^8B$ solar neutrinos on Deuterium through

$$\nu_e~+~D~\rightarrow~p~+~p~+e^-\ ,$$
with physics results soon to come. These results will address the distortion of the spectrum as well, but at this early stage of data acquisition, comparison with the SuperK results might prove more telling.  Since SuperK detects neutrinos through elastic scattering of electrons, a process mediated both by charged and neutral currents, comparison will shed light on the amount of neutral current. Since neutral current interactions are flavor blind, it should say something about the nature of the neutrino into which $\nu_e$ oscillates.  Eventually, neutron catchers will be installed in the SNO tank, which will allow for the detection of the neutral current reaction

$$\nu_e~+~D~\rightarrow~p~+~n~+\nu_e\ ,$$
and test directly the idea of flavor oscillations. 

\item The KAMLAND experiment\cite{KAMLAND} will soon be able to distinguish between large and small mixing angle solutions.

\item  MiniBoone\cite{MINIBOONE} at FermiLab will address the LSND result and determine if a fourth neutrino is needed.

\item BOREXINO\cite{BOREXINO} in the Gran Sasso tunnel in Italy will measure the amount of solar neutrinos produced in Beryllium capture.

\item MINOS\cite{MINOS} from FermiLab to the Soudan mine will monitor a muon neutrino beam over long distances.

\item GNO\cite{GNO}at Gran Sasso continues to measure the low energy solar neutrino spectrum.
\end{itemize}
In the next five years, new data should decide three burning issues: the need for a sterile neutrino,  the mixing angle for the electron neutrino (large or small?), and if the solar neutrino deficit is due to oscillations.

We do not know if the Standard Model symmetries exhaust all the symmetries of the brane on which we live. It could easily be that the brane picture is valid up to Planck or GUT scale, as there are no compelling reasons for the brane description to break down at the TeV scale. If there are more symmetries on our brane, these could be in the form of GUT symmetries, or else simpler Abelian remnants which somehow are reflected in the masses and mixings of the elmentary particles. 
Generic $SO(10)$  requires one sterile neutrino per family, and its naive application yield large mixing between $\nu_\mu$ and $\nu_\tau$\cite{HRR,CASE}, an observed feature at SuperK, and small mixings between $\nu_e$ and $\nu_\mu$. The latter seems disfavored by the recent SuperK solar neutrino results. Some models of the Froggatt-Nielsen type\cite{FN} attribute the observed hierarchies to Abelian symmetries; these are models of Cabibbo suppression, and they also tend to yield unsuppressed $\nu_\mu-\nu_\tau$ mixing and highly suppressed $\nu_e-\nu_\mu$ mixing~\cite{EIR}. Should the mixing turn out to be large, these models will lose some of their appeal. On the other hand, there does not seem to be any preferred mixing scheme coming from the brane world picture of neutrino masses\cite{ADDM,OTWN}, where most models are in their postdictive phase.  

As we learn more about the neutrino world, we may hope  that the new measurements will point to  particular physics beyond the Standard Model. As they have in the past, neutrinos seem to contain many clues; it is up to us to recognize them.

\section*{Ack$\nu$ledgments}
I wish to thank Professors P. Frampton and J. Ng for their kind invitation to speak in such civilized surroundings. I also wish to acknowledge support from the Department of Energy 
under Grant DoE DE-FG02-97ER41029.

\end{document}